\documentclass{aa} 

\def\ergscm2{\rm erg\,cm^{-2}\,s^{-1}}
\def\ergs{\rm erg\,s^{-1}} 
\def\cm2{\rm cm^{-2}} 
\def\scm2{\rm cm^{-2}\,s^{-1}}
 
\def\BSAX{{\em Beppo}SAX }
\def\exo{EXO 053109-6609.2}

\input psfig.sty 

\begin{document} 
 
\title{Timing and spectral changes of the Be X-ray transient EXO\,0531-6609.2 
        through high and low state.}
 
\author{Nanda Rea $^{1,2}$, Gian Luca Israel $^{2,3}$, Tiziana Di Salvo$^{4,5}$, Luciano Burderi$^2$, Gabriele Cocozza $^1$}

\institute{ 
University of Rome 2, Via della Ricerca Scientifica 1,
I--00133 Roma, Italy: rea@mporzio.astro.it
\and
INAF--Astronomical Observatory of Rome, Via Frascati 33,  
I--00040 Monteporzio Catone, Italy
\and
Affiliated with ICRA
\and
Astronomical Institute "Anton Pannekoek", University of 
 Amsterdam and Center for High-Energy Astrophysics,
 Kruislaan 403, NL 1098 SJ Amsterdam, the Netherlands
\and
Dipartimento di Scienze Fisiche ed Astronomiche, Universita' di Palermo,
via Archirafi 36, 90123 Palermo, Italy 
}

\date{  30 Jan 2004 } 
\offprints{rea@mporzio.astro.it} 
\authorrunning{Rea et al.} 
\titlerunning{\exo\, through high and low state} 

\abstract { We report on spectral and timing analysis of \BSAX data of
the 13.6\,s period transient X-ray pulsar \exo. Observations were
carried out in March 1997 and October 1998, catching the source during
a high and a low emission state, respectively. Correspondingly, the
X-ray luminosity is found at a level of $ 4.2 \times 10^{37} \ergs $
and $1.5\times 10^{36} \ergs $ in the two states.  In the high state
the X-ray emission in the energy range 1--100\,keV is well fitted by
an absorbed power--law with photon index $ \Gamma\sim 1.7$ plus a
blackbody component with a characteristic temperature of $\sim
3.5 $\,keV. Moreover, we find an evidence of an iron emission at
$\sim$6.8\,keV, typical feature in this class of sources but never
revealed before in the \exo\, spectrum. In the low state an
absorbed power--law with $\Gamma\sim 0.4$ is sufficient to fit the
1--10\,keV data.  During \BSAX observations \exo\, display
variations of the pulse profile with the X-ray flux: it showed single peaked and
double peaked profiles in the low and high state, respectively.  Based
on these two observations we infer a spin--up period derivative of
$- (1.14\pm0.08)\times 10^{-10}ss^{-1}$. By comparing these with other
period measurements reported in literature we find an alternating
spin-up and spin-down behaviour that correlates well with the X-ray
luminosity.
 }

\maketitle
 
\section{Introduction} 

\exo\, is a High Mass X-ray Binary (HMXB) hosting a neutron star and a
B--emission spectral type (Be) star (Haberl, Dennerl, \& Pietsch 1995;
McGowan \& Charles 2002).  The system is in the Large Magellanic Cloud
(LMC), $\sim 17^{\prime}$ away from LMC X-4, another high mass X-ray binary
pulsar (La Barbera et al. 2001).  Be stars are characterised by high rotational
velocities (up to 70\% of their break--up velocity), and by episodes
of equatorial mass loss which produce a temporary ``decretion'' disk
around the star (Slettebak 1987, Okazaki \& Negueruela 2001). These
X-ray binaries in which a neutron star is orbiting in a relatively
wide orbit with moderate eccentricity are the most numerous among
HMXBs. In the LMC, more than half of the confirmed HMXBs are variable
or transient sources consisting of a neutron star with a Be companion
(Haberl, Dennerl, and Pietsch 2003; Sasaki, Haberl \& Pietsch 2000;
Haberl \& Sasaki 2000).

Variable X-ray emission is often observed which likely results from
the large variations in the Be wind density and relative velocities
along the neutron star orbit, which in most cases may results in
regular X-ray outbursts near the periastron (Type I outbursts).
Alternatively, aperiodic outbursts occur which often last longer
than the neutron star orbit. These are probably caused by matter
ejection outflowing from the equatorial plane of the Be star (Type II
outbursts; Stella, White \& Rosner 1986; Motch et al. 1991).

\exo\, was discovered in 1983 deep EXOSAT exposures of the LMC X-4
region (Pakull et al. 1985; Pietsch, Dennerl, and Rosso 1989). The
luminosity was $\sim 6 \times 10^{36} \ergs $ (0.15--4 keV; assuming a
distance of 50 kpc). Subsequently, the coded mask X-ray telescope SL2
XRT flown on board of the shuttle Challenger between July and August
1985 detected a second outburst (Hanson et al. 1989), with a source
luminosity of $\sim 1 \times 10^{37} \ergs $ (2--10\,keV).

The source was then monitored from June 1990 to July 1994 with the
ROSAT PSPC (Haberl, Dennerl, \& Pietsch 1995). Alternating high states
($L_{x}\sim 10^{37}\ergs $) and low states ($L_{x}\sim 10^{36}\ergs $) in the
X-ray flux were discovered.  Haberl et al. 1996 detected another
outburst from March to May 1993 with an average luminosity $\sim
2.4 \times 10^{36} \ergs $ (0.1--2.4\,keV).

Dennerl et al. (1996) reported on a ROSAT observation that took place
in October 1991 and led to the detection of a spin period of
13.67133(5)\,s with a period derivative of $ (1.5\pm0.1)\times10^{-8}
ss^{-1}$ (calculated during their observation).  Haberl et al. 1995
proposed an orbital period of about 600--700 days and an orbital
eccentricity of $e \sim0.4-0.5$.  Under the assumption
that period changes are caused by Doppler shifts, Dennerl et
al. (1996) corrected the proposed orbital period, finding an orbital
solution with $ P_{orb} = 25.4 $ days and an eccentricity of $ e \sim
0.1 $.

Timing analysis of the \BSAX\ data obtained in March 1997 when the
system was in high--state, revealed coherent pulsations at a period of
13.67590(8)s and two different period derivatives: a short-term period
derivative, calculated during the 2 days of BeppoSAX observation, of
$\dot{P}_{loc}=(3.7\pm0.5)\times10^{-9} ss^{-1}$, and a secular period
derivative, calculated from a comparison with a previous ROSAT measure
of $\dot{P}_{sec}=(3.67\pm0.05)\times10^{-11} ss^{-1}$ (Burderi et
al. 1998).

Last observation of \exo\, was a deep XMM-Newton observation of LMC
field of October 2000 (Haberl, Dennerl and Pietsch 2003) in which
\exo\, was found to be the brightest source in the field
($7\times10^{-12}\ergscm2$ in 0.2-10\,keV range, corresponding to a
luminosity of about $2.1\times10^{36}\ergs$).

Across different observations this source varied its intensity up to
a factor of about 10. Nevertheless, notwithstanding most of Be X--ray
binary systems show a quiescent state, this source has not yet been
detected in quiescence but only in a low--luminosity state with
a minimum luminosity of $\sim 10^{35}-10^{36}\ergs$.

In this paper we report on spectral, spin period and pulse profile
changes with the X-ray luminosity. We finally compare our results
with previous findings.

%%%%%%%%%%%%%%%%%%%%%%%%%%%%%%%%%%%%%%

\begin{figure}[htb] 
\centerline{\psfig{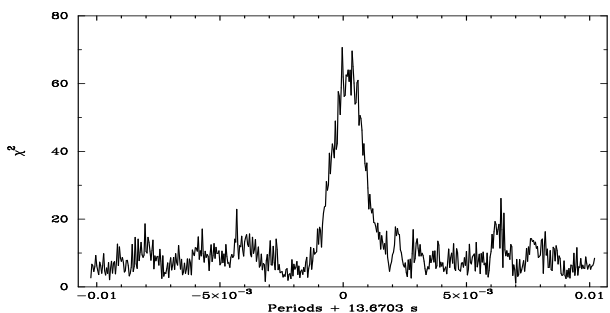} } 
\caption{Epoch folding search of the data from the low state observation of 1998.} 
\end{figure} 

\begin{figure}[htb] 
\centerline{\psfig{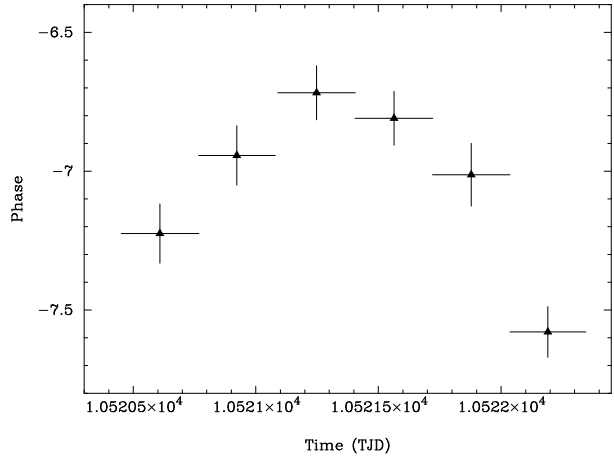}} 
\caption{Evidence of a quadratic component in the phase--fitting of the 
high--state observation (linear component, corrisponding to the
local $\dot{P}$ was removed in order to better show the quadratic
residuals).}
\end{figure} 

%%%%%%%%%%%%%%%%%%%%%%%%%%%%%%%%%%%%%%

\section{Observations}

\BSAX\, observed \exo\, with its narrow-field instruments (NFI; Boella et al. 1997a) 
from 10:48:16 of 1997 March 13 to 08:17:50 of 1997 March 15 and from
22:39:47 of 1998 October 20 to 08:05:31 of 1998 October 22.

%%%%%%%%%%%%%%%%%%%%%%%%%%%%%%%%%%%%%%

\begin{figure}[!ht] 
\vbox{ 
\centerline{\psfig{figure=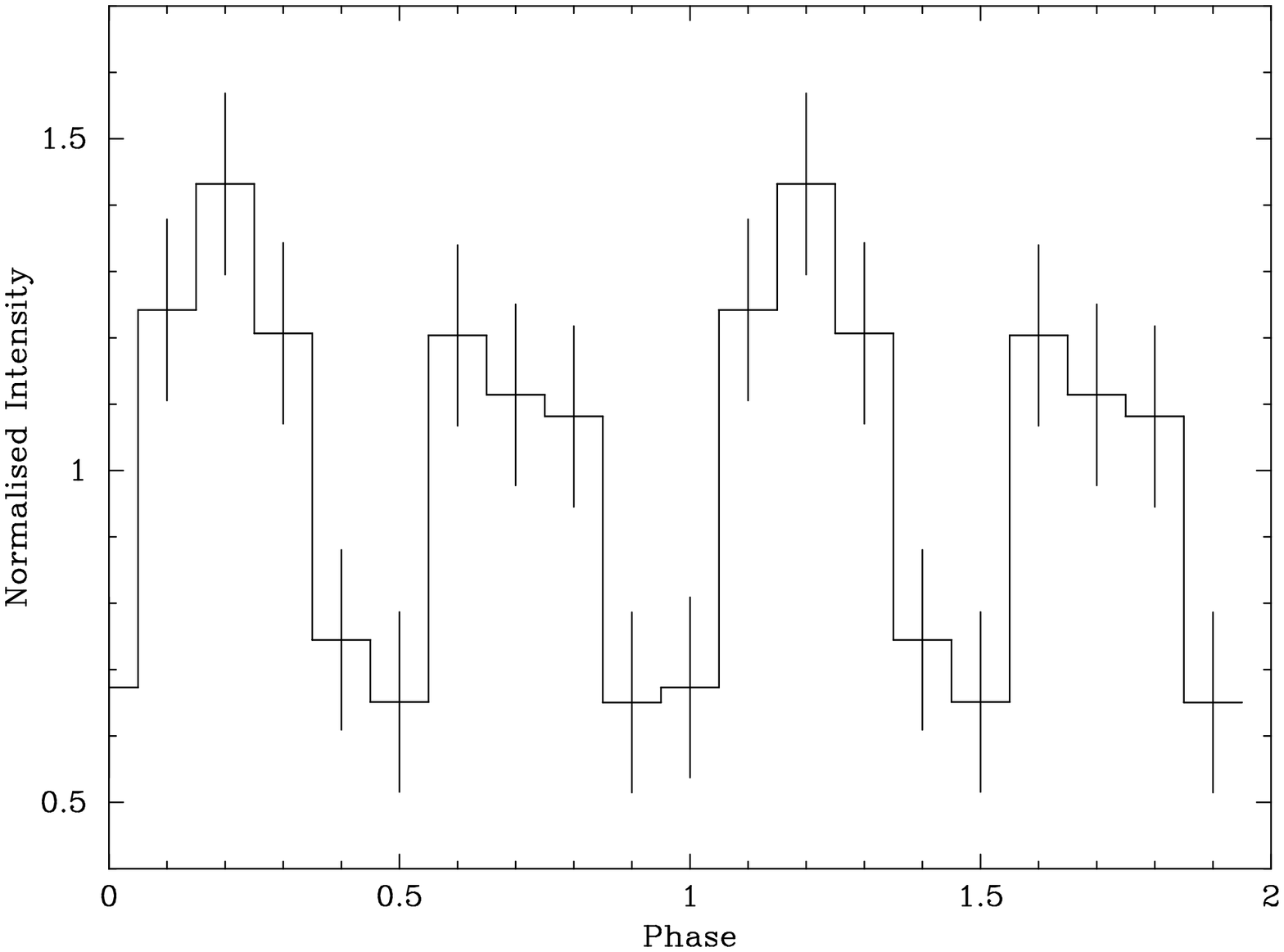,width=8cm,height=6.5cm,angle=270}}
\centerline{\psfig{figure=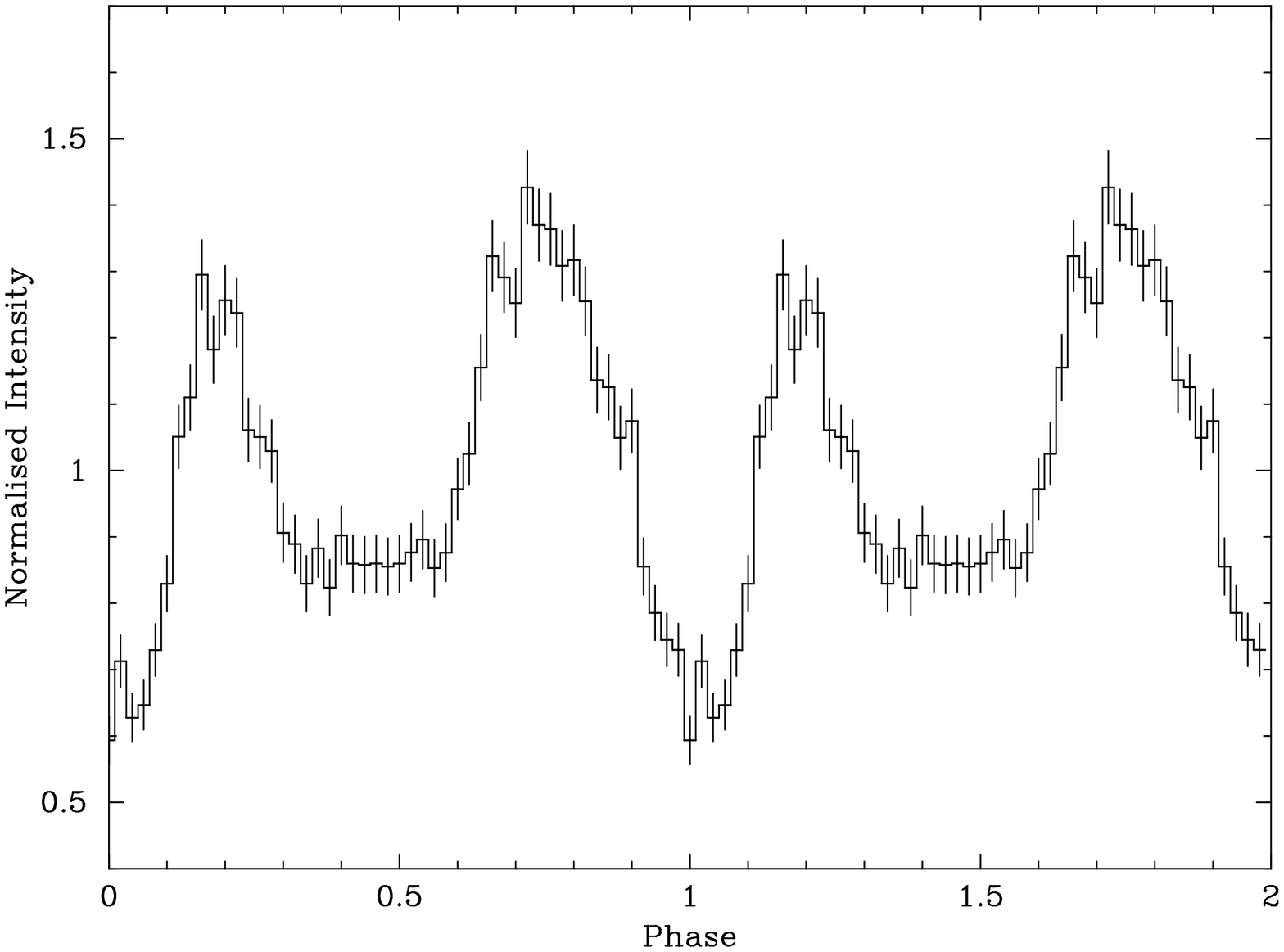,width=8cm,height=6.5cm,angle=270}}
\centerline{\psfig{figure=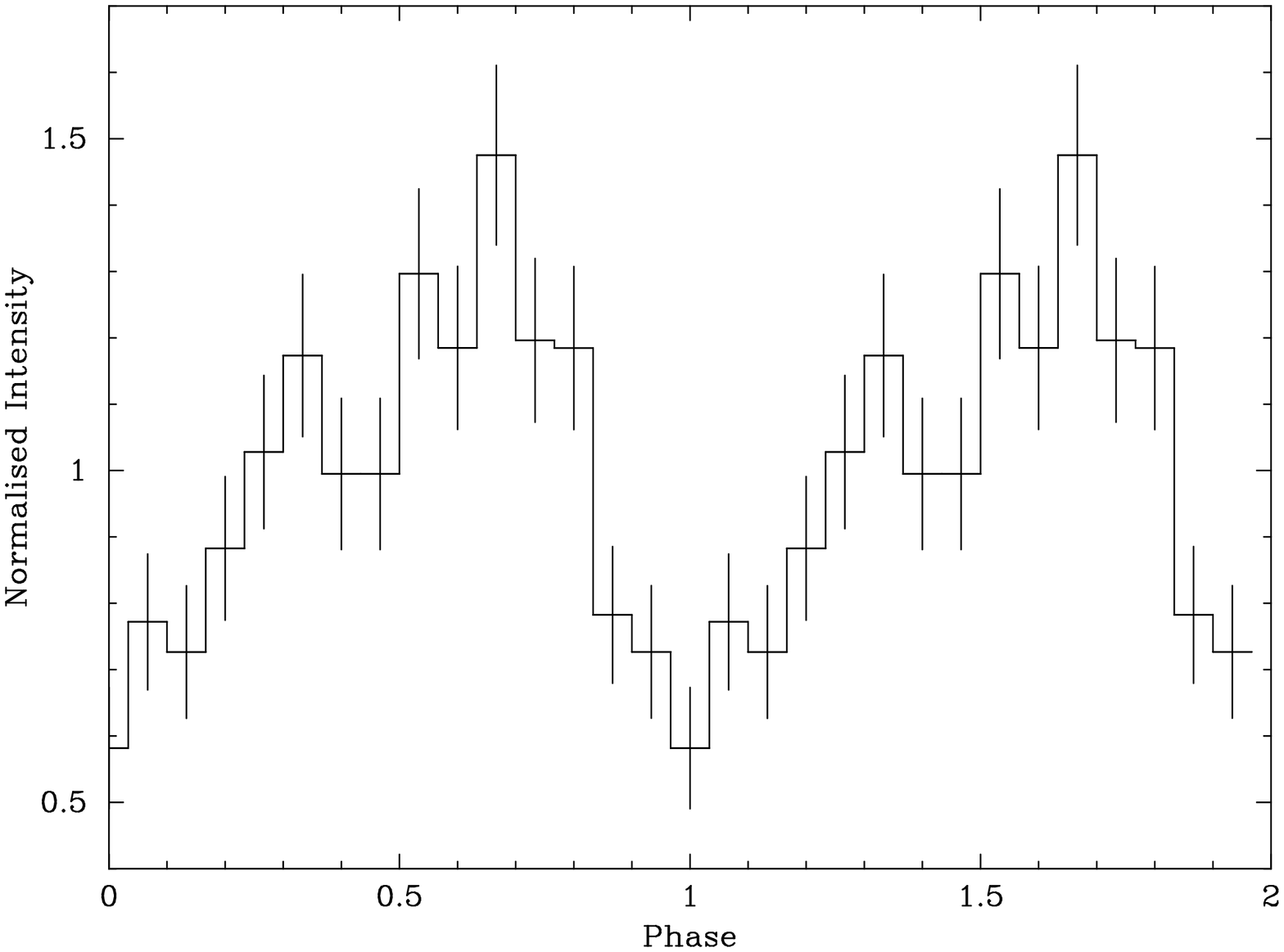,width=8cm,height=6.5cm,angle=270} } 
\caption{First and second panels: pulse profile of March 1997 \BSAX\,observation 
of \exo\, in PDS and MECS bands aligned in phase, respectively. Third panel: pulse profile of 
October 1998 \BSAX\,observation of \exo\, in MECS band, while the source was 
in low--state (not phase aligned with previous observation) .}
}
\end{figure}

%%%%%%%%%%%%%%%%%%%%%%%%%%%%%%%%%%%%%%%

Spectral coverage of \BSAX\, extends over a wide energy range using 
four different instruments. In our analysis we use just two coaligned
instruments: Medium-Energy Concentrator Spectrometer (MECS) consisting
of three units operating in the 1-10\,keV range (Boella et al. 1997b) and
the Phoswich Detector System (PDS) with four scintillation units
operating in the range 10-200\,keV (Frontera et a. 1997).  The High
Pressure Gas Scintillator Proportional Counter (HPGSPC, 7--40\,keV;
Manzo et al. 1997) was not available; and for what concerns the Low
-Energy Concentrator Spectrometer (LECS, 0.1--4\,keV; Parmar et
al. 1997), in the low state observation there are not enough photons
to be used, while for the high state the off--axis response matrix is
not accurate enough (actually LECS data were not fully consistent with
the MECS data in the overlapping range $\sim$1--5\,keV).  Both
instruments used have a field of view less than one degree but only
MECS has imaging capabilities. 

We extracted \exo\, data from $ 4'$ radius circular regions in the
MECS field of view, centred on the maximum of the point spread
function of the source. The data were background subtracted by
extracting in the same field of view background photons from a similar
circular region of $4'$ radius, centred far enough from any sources
present in the image and at the same off--axis position of \exo; the
arrival times were then corrected for the barycentre of the solar
system. Owing to the problems descrived above, we used only MECS and
PDS data for the first observation: we could use the PDS data because
the contaminating source LMC X-4 was in a low state during that
observation (Burderi et al. 1998). On the other hand, during the
second observation LMC X-4 was in a high state (Naik \& Paul 2003). In
the PDS range, the relative fluxes between \exo\, and LMC X--4 are:
$F_{EXO}/F_{LMC}\sim 9$ for the first observation, and
$F_{EXO}/F_{LMC}\sim 0.2$ for the second (latter value was inferred
extrapolating \exo\, low state spectrum in the PDS range).  
Therefore, because during the second observation LMC X-4 was greatly
more luminous than \exo,  and given that PDS does not have imaging 
capabilities, we only used MECS data in this case. 

MECS spectra, accumulated from the same circular regions used for the
event files, were re-binned with the \BSAX\, grouping file:
mecs\_5.grouping; PDS spectra were re-binned in order to have at least
fifty photons for each bin.  Obviously, both re-binning are made such
that minimum chisquare techniques could be reliably used in spectral
fitting.  Spectral analysis was performed using the energy range:
$1.65-10.5$\,keV for the MECS and $15-60$\,keV for the PDS (data in
the 60--100\,keV PDS range was bad), and taking into account the
off-axis position of the source. We used off--axis response matrices
made using the precise estimation of \exo\, off axis.
Note that by using on--axis spectral response matrices for a
source that is off--axis, one wrongly estimates the source spectral
parameters. In particular, the flux is underestimated while the
spectrum appears softer than what it actually is (Mereghetti et
al. 2000). Correspondingly, any extrapolation towards high--energies
may be strongly underestimated. All energy bins the count rate of
which was consistent with zero were not used in the spectral analysis.

\section{Timing analysis}

We search for coherent pulsations of the source in both states. In the
low state (second observation), in a period interval around the ROSAT
period, the power spectrum and the epoch folding search show an
evident signal at $\nu \sim 0.0732$ Hz with a high significant peak
$(\sim10\sigma)$ corresponding to a period of $\sim 13.67$\,s (see
Fig.1). Hereafter, if not specified, all errors are 1$\sigma$
confidence level. To rely upon a more accurate period measurement, we
used a phase fitting technique by fitting the phases of the modulation
over four intervals of $\sim$ 21000\,s each. With this method we found
a spin period of $13.6705\pm0.0004$\,s.

Timing analysis of the first observation was already performed giving
a spin period value of 13.67590(8)\,s (Burderi et al. 1998).  Applying
the same method to the March 1997 observation, we found a spin period
of $13.67630\pm0.00003$\,s (consistent with the period found for the
same observation in Burderi et al. 1998)\footnote{Note that while the
period in Burderi et al. 1998 is referred to the start of the
observation, here we report the periods at the half time of the
observation. Taking in account the local period derivative reported
before, the two periods are consistent.}. In the latter case, owing to
a higher statistic, we looked at the phase modulations over six
intervals of $\sim$ 14000\,s: fitting the results with a polynomial
function of the second order we found evidence for a quadratic
component possibly related to the orbital motion of the X--ray pulsar
(see Fig.2).  The observation is unfortunately too short to look for
an orbital solution.  From the periods measured in these two
\BSAX\,observations we inferred a $\dot{P} = -(1.14\pm0.08)\times
10^{-10} ss^{-1}$.

The pulse profile obtained when the source was in high--state (March
1997) is dominated by the first harmonic. This predominance disappear
at fainter fluxes (October 1998) where the fundamental is the
strongest component (Fig.3).  Furthermore, the pulsed fraction (PF) in
the MECS range changes from a value of $50\pm1$\% to $30\pm 1$\% from
high to low state, while in the PDS range only in the high state we
see the pulsation, and the PF is $40\pm4$\% (see Fig.3). The low state
PDS epoch folding at the period found in the MECS data did not show
any modulation due to poor statistics (with an upper limit on the PF
is $\sim$ 80\% or 100\% depending if we estimate the PDS flux of LMC
X-4 be 3.5 counts/s or 4.0 counts/s), correspondingly we cannot
exclude an high-energy pulsation of the source in low state.

%%%%%%%%%%%%%%%%%%%%%%%%%%%%%%%%%%%%%%%%

\begin{table}[bht] 
\caption{Spectral parameters of EXO 0531-66 in high--state (absorbed 
blackbody plus a power--law and a Gaussian) and in low--state (absorbed power--law). 
Fluxes are unabsorbed and in unit of 
$\times 10^{-11}\ergscm2$ (see also Fig.4); errors are 1$\sigma$ 
confidence level.} 
\begin{tabular}{lcc} 
\hline  

Spectral parameters          &  High state & Low state \\
\hline  
 
$N_{H} (\times 10^{22}\cm2)$  &  $0.92\pm0.04$ &  0.92 (frozen)   \\ 

Power--law PI &          $ 1.75\pm0.08 $   &   $0.4\pm0.1$    \\

flux PL (2--10 keV) & 2.0 &  -- \\ 

$kT_{bb}$\,(keV)                &3.53  $\pm0.3$ & --  \\

flux bb (2--10 keV)  &   1.9& -- \\  

$E_{Fe}$ (keV) & $6.8\pm0.1$ & -- \\

$\sigma_{Fe}$ (keV) & $0.4\pm0.1$ & -- \\

$EQW_{Fe}$ (eV) & 275 & -- \\

$ {\chi_{\nu}}^2 $ &   1.18 (169 dof) &   1.2 (25 dof) \\

Total Flux (2--10 keV)     & 5.0  & 0.5 \\
Total Flux (2--60 keV)    & 14.0  & -- \\

\hline  
\end{tabular}   
\label{tab: lines} 
\end{table}

%%%%%%%%%%%%%%%%%%%%%%%%%%%%%%%%%%%%%%%%

\begin{figure}[!ht]
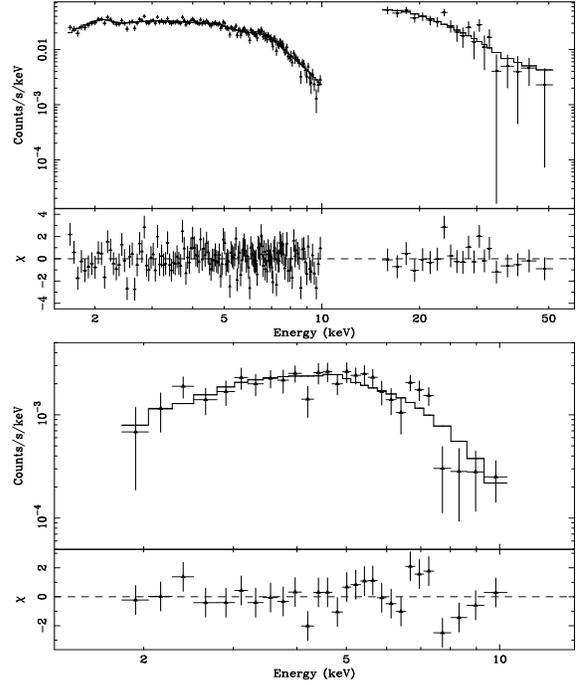
 
\vbox{
\centerline{\psfig{figure=fig6.ps,width=7.5cm,height=4.5cm,angle=270}} 
\centerline{\psfig{figure=fig7.ps,width=7.5cm,height=4.5cm,angle=270}} 
\caption{\BSAX\, spectra of \exo\, in high state (1997) and in low state (1998).
 See Tab.1 for spectral parameters.} 
}
\end{figure}

%%%%%%%%%%%%%%%%%%%%%%%%%%%%%%%%%%%%%%%%

\section{Spectral analysis}

\exo\, high--state spectrum is well fitted by an absorbed blackbody
plus power--law component and a Gaussian function (see Tab.1 and
Fig.4). The inferred luminosity between 2--60\,keV, assuming a
distance of 50 kpc (this is the distance known to the Large
Magellanic Cloud), is $4.2\times10^{37}\ergs$.  We tried to fit the
spectrum with other models: two blackbody plus a power--law, a
blackbody plus a cutoff power--law and a blackbody plus a
high--energy cutoff model but all gives worst results ($\chi_{\nu}^2 >
1.9$).  Actually, fitting the spectrum with a blackbody plus a cutoff
power--law we initially obtained a good chisquare value but the
absorption was too small ( $N_{H}=(0.03\pm0.01)\times10^{22}$
atoms/$cm^2$) for a system in the LMC. We then fixed the absorption
value at $\sim 0.1\times10^{22}$ atoms/$cm^2$ but the chisquare becomes
remarkably worse (this is the galactic absorption value calculated for
the RA and DEC of \exo\, with the {\em nh} ftool; Dickey \& Lockman,
1990).  Adding a Gaussian line at $\sim$ 6.8\,keV (see Tab.\,1) to the
continuum model the chisquare value is improved from 1.28 (172 dof)
to 1.18 (169), giving an F test probability of $3.5\times10^{-4}$.

The low--state spectrum is modelled by an absorbed power--law (see
Tab.1 and Fig.4) and the source luminosity in this state is $
1.5\times10^{36}\ergs$ (for a distance of 50\,kpc; in 2-60 keV
energy range). We also tried to add a Gaussian function at $\sim$ 6.8\,keV 
but it was no significative (confidence level $< 1.5\sigma$).

\section{Discussion}

\begin{figure*}[htb] 
\centerline{\psfig{figure=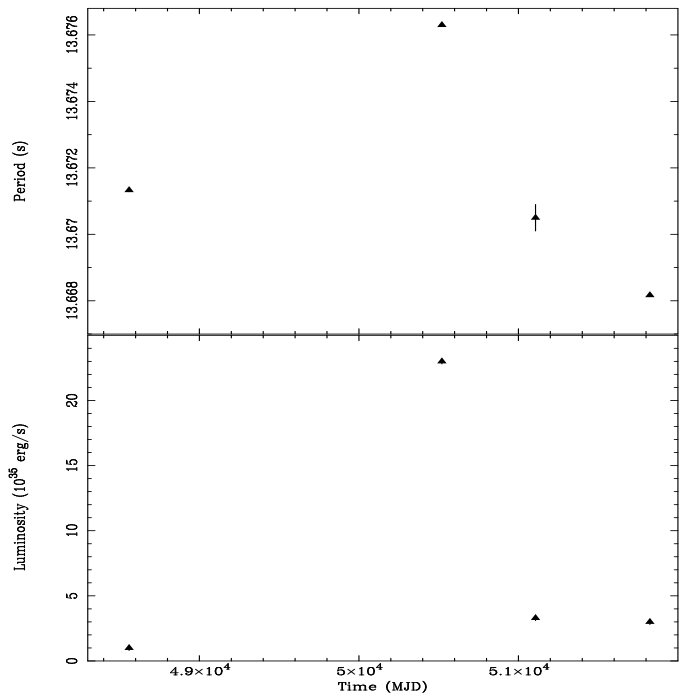,width=10cm,height=10cm}} 
\caption{Spin period and luminosity changes between ROSAT 
(Haberl et al. 1996), \BSAX\, (this paper) and XMM (Haberl et al. 2003) 
 observations (all reported luminosities are extrapolated in 0.1--2.4\,keV band).}
\end{figure*} 

During the outburst phase of Be X--ray binaries, accretion disks are
expected to be present, and indeed, evidence for an accretion disk,
based on a correlation between the observed flux and spin--up or
spin--down rate , has been found for several sources (Bildsten et
al. 1997; Okazaki \& Negueruela 2001).  The \exo\, spin period was
detected four times: the first time by ROSAT (Dennerl et al. 1996),
two times by \BSAX\, (Burderi et al. 1998 and this paper) and the
fourth time by XMM-Newton satellite (Haberl et al. 2003).  In Fig.5 we
plot the 0.1--2.4\,keV band luminosity versus period for all these
four observations; it is evident from the plot that periods and
luminosities are directly correlated.  Comparing the spin periods of
all the observations of \exo\, carried out from its discovery, we
found that the source alternates spin--up and spin--down (Fig.5).  The
secular period derivative between the ROSAT observation (Haberl,
Dennerl, and Pietsch 1995) and the first \BSAX\, observation is
$(2.9\pm0.1)\times 10^{-11}ss^{-1}$; comparing the two \BSAX\,
observations we obtain $-(1.14\pm0.08)\times 10^{-10}ss^{-1}$, and the
$\dot{P}$ between the last \BSAX\, observation and the XMM observation
(Haberl et al. 2003) is $-(3.7\pm0.1)\times 10^{-11}ss^{-1}$ (period
derivative errors are at 90\% confidence level).  The secular spin
period derivative is changed from a spin--down to a spin--up trend,
which has a clear correlation with the X-ray luminosity. We detected
this change around 1997.

Double peaked pulse profile (Fig.3, first and second panels) was
observed when the source was in high--state, while a single peaked
profile was found in the low--state (Fig.3, third panel). This
behaviour has been often seen in X--ray binary systems and has been
ascribed to the transition of the source between a pencil beam and a
pencil plus fan beam emission geometry, a behaviour which is
correlated with the X--ray flux (Parmar et al. 1989a and
1989b). Timing behaviour of this source is similar to what seen in
other Be X--ray binaries, making the period derivative changes a
peculiar characteristic of these systems.

The spectrum in the high--state of emission is well fitted by an
absorbed blackbody plus a power law and a Gaussian line at $\sim$
6.8\,keV (see Tab.1 and Fig.4 first panel).  
The 6.8 keV emission line is probably due to K-shell emission from highly
ionized iron (probably in the He-like and/or H-like ionization stages).
This is the first evidence of an iron emission line in this source; iron
emission lines are a common feature in HMXBs, although their centroid
energies are usually detected in the range 6.4 - 6.7 keV (e.g. Parmar et 
al. 1989a).

From measured temperature of kT$\sim3.5$ keV we infer an emission
radius for the blackbody component of $\sim 5.5$ km, about half
neutron star radius.  The calculated blackbody radius is smaller than
the neutron star radius, probably due to the fact that the thermal
emission does not come from all neutron star surface but probably just
from a small region around the polar caps. The energy flux of the two
spectral components differs of about two orders of magnitude,
$F_{bb}/F_{power} = 1.728\times10^{-2}$.  The blackbody component
probably disappear while the system is in the low--state and the power--law
photon index becomes flatter (see Tab.1 and Fig.4 second panel).

\vspace{1cm}
We thank T.Mineo for useful helps with LECS response matrices. 
This work is born during an Astrophysical Laboratory of
Prof. R.Buonanno in the University of Rome ``Tor Vergata''. This work
is supported through ASI, CNR and Ministero dell'Universit\`a e
Ricerca Scientifica e Tecnologica (MURST-COFIN) grants.

\end{document}